\documentstyle[12pt,aps]{revtex}
\def\ii{\'{\i}}
\def\bb{$\beta\beta_{2\nu}$~}
\def\ru{$^{100}Ru$}
\def\mo{$^{100}Mo$}
\begin{document}
\title{Double beta decay of $^{100}Mo$: the deformed limit}
\author{Jorge G. Hirsch\thanks{e-mail: hirsch@fis.cinvestav.mx}\\
{\tenit Departamento de F\'{\i}sica, Centro de Investigaci\'on y de
Estudios Avanzados del I. P. N.,\\ A. P. 14-740 M\'exico 07000 D.F.}\\
O. Casta\~nos\thanks{e-mail:ocasta@redvax1.dgsca.unam.mx} and
P. O. Hess\thanks{e-mail:hess@xochitl.nuclecu.unam.mx}\\
{\tenit Instituto de Ciencias Nucleares, Universidad Nacional Aut\'onoma de
M\'exico,\\ A. P. 70-543 M\'exico 04510 D.F. }\\
Osvaldo Civitarese\thanks{Fellow of the CONICET,
Argentina; e-mail:civitare@venus.fisica.unlp.edu.ar}\\
{\tenit Departamento de F\ii sica, Universidad Nacional de La Plata,\\
c.c.67; 1900, La Plata, Argentina}   }
\maketitle

\vskip 1cm
\abstract{The double beta decay of $^{100}Mo$ to the ground state and
excited states of $^{100}Ru$ is analysed in the context of the pseudo
SU(3) scheme. The results of this deformed limit are compared with the
vibrational one based on the QRPA formalism. Consistency  between the
deformed limit
and the experimental information  is found for various $\beta\beta$
transitions, although, in this approximation
some energies and B(E2) intensities cannot reproduced.}


\vskip 1cm

The two neutrino mode of the double beta decay ($\beta\beta_{2\nu}$) is
an exotic but allowed second order process in the standard model. It has
been detected in nine nuclei and has served as a tool to develop
the most refined experimental detection methods for very low signals.
It also represents a severe test on
the nuclear structure description of those nuclei\cite{Doi85}.

The $^{100}Mo$ is an excellent subject of study. Its large Q-value ,
$Q^{\beta\beta} = 3.034 MeV$, favors its decay. Four groups have
reported on
the detection
of the $\beta\beta_{2\nu}$ of  $^{100}Mo$ to the ground state of
$^{100}Ru$ with half-lives of the order of
$10^{19}$yrs\cite{Eji91,Als93,Moe93,Nem94}.
Also the decay to the first excited $0^+_1$ state has been
reported\cite{Moe93,Bar91} .

This abundance of experimental information allowed a careful check of
the nuclear models used to describe the two neutrino mode. The QRPA
model, both in its conventional and number projected versions, was found
to collapse when realistic particle-particle interactions are
used\cite{Vog86,Civ91}.
Under the assumption that the excited 1130 keV $0^+$ state in $^{100}Ru$
is a member of a $0^+, 2^+, 4^+$ two-phonon vibrational triplet
the QRPA calculation exhibits an overestimation of the	amplitude of the
$\beta\beta_{2\nu}$ decay to this excited state. However the decay to the
ground state is reasonably well reproduced by using the
same approximation.
The single beta-decay
transition between the initial and intermediate nuclei is also overestimated
by the theory\cite{Gri92}. Assuming realistic nucleon-nucleon
interactions similar results were found, {\em i.e.} the
$\beta\beta_{2\nu}$ decay amplitudes are overestimated. This result is
particularly true for the decay to the first $2^+$ state\cite{Suh94}.

All the works dealing with the \bb ~decay of $^{100}Mo$ found a
strong dominance of one single particle transition: $(g_{7/2}(n))^2
\rightarrow (g_{9/2}(p))^2$.
This lack of collectivity was mentioned as a
possible cause of the failure of
the QRPA to describe the \bb ~decay of this specific
nucleus\cite{Civ91}  while having success in
others\cite{Vog86,Civ87,Mut89,Hir90}.

Other complementary explanation could be that  deformation effects
were destroying the
one-phonon--two-phonon structure in $^{100}Ru$, and substituting it
with rotational bands built upon a deformed ground state or some
vibrational intrinsic excitations\cite{Suh94}.
In the present paper we explore both possibilities using the pseudo
SU(3) formalism.

The transition from spherical to deformed shapes in $Mo$ isotopes was
study microscopically several years ago, commenting on the
deformed character of some first excited $0^+$ states\cite{Fed78}. In
a similar way within
the General Collective Model (GCM)\cite{Hes80} a study of the potential
energy surfaces (PES) of the $Ru$ isotopes\cite{Tro91} exhibited a
transition from
a harmonic oscillator ($^{98}Ru$) to a nucleus with triaxial dominance
($^{108}Ru$).
All $Ru$ isotopes between $^{100}Ru$ and $^{104}Ru$ behave very
similar\cite{Tro91}. The PES has a global minimum
at $\beta =0$, suggesting an harmonic oscillator for the
low lying states. There is also a local minimum at $\beta$
approximately $0.4$ and $\gamma =24^o$, which suggests triaxial shapes
for excited states.
However the energy of the ground state lies {\em above} the potential
barrier between the spherical and the triaxial minimum. As a consequence, the
ground state, the first and second excited $2^+$
state are spread out between the global and triaxial minimum showing the
behavior of an anharmonic oscillator, while the first excited
$0^+$ and third excited $2^+$ state are peaked within the triaxial
minimum. As a picture we obtain the spectrum of an anharmonic oscillator
were the first excited $0^+$ state does not fit into it.
Nevertheless the $B(E2)$ transitions follow the characteristic of
an anharmonic oscillator within a simple consideration, i.e., the transition
from the first exited $0^+$ state to the first excited $2^+$ state is
strong compared to the second excited $2^+$ state, which is weak.
This did lead in the past to the assumption that $^{100}Ru$ resembles
more an anharmonic oscillator.

 We will use the pseudo SU(3) model\cite{Dra84} to provide a
microscopical description of the ground and excited low energy states in
\ru , and the ground state of \mo  ~including deformation. This exercise
will supply a rotational limit in the description of these nuclei, to be
compared with the spherical, vibrational limit obtained with the QRPA.

The double beta decay, when described in the pseudo SU(3) scheme, is
strongly dependent on the occupation numbers for protons and neutrons in
the normal and abnormal parity states $n^N_\pi, n^N_\nu, n^A_\pi,
n^A_\nu$\cite{Cas94}. In particular, the \bb decays are allowed only if
they fulfil the following relations:

\begin{equation}
\begin{array}{l}
n^A_{\pi ,f} = n^A_{\pi ,i} + 2,
\hspace{1cm}n^A_{\nu ,f} = n^A_{\nu ,i}
\hspace{1cm}n^N_{\pi ,f} = n^N_{\pi ,i} ,
\hspace{1cm} n^N_{\nu ,f} = n^N_{\nu ,i} - 2
\end{array}\label{num}
\end{equation}

These numbers are determined by filling the Nilsson levels from below,
as discussed in \cite{Cas94}. They are exhibited in Table 1 as the {\em
deformed occupation (def-o)} case. Both \mo  ~and
\ru ~have deformations $\beta \approx 0.23$. This number must be taken
carefully, given these nuclei do not exhibit a rotational spectra, but
is a hint about their {\em average} deformations.
We were forced to select a slightly higher deformation for neutrons than
for protons. If not, the removed pair of neutrons in \ru ~would come from
the abnormal parity orbital, and in this case the expressions (\ref{num})
would not be fulfilled suppressing the \bb decay.

In the third and fourth row of Table 1	the occupations in the
{\em spherical occupation (sph-o)} limit ($\beta \leq 0.09$) are given.
This is a very interesting
limiting case, in which all the proton holes belong to the
abnormal $g_{9/2}$ orbit and all the neutron particles to the normal parity
orbitals.

In the abnormal parity space only seniority zero
configurations are taken into account. This is a very
strong assumption, which in future works is expected to be improved,
but is quite useful in order to simplify the calculations.
Its effects in the present calculation are discussed below.

Under the assumption of a Hamiltonian consisting in a Nilsson
mean field and a strong
quadrupole-quadrupole interaction, the deformed wave
functions of \mo  ~and \ru , with angular momentum $J$ and belonging to
the ground state ($\sigma = g.s.$) or the excited ($\sigma = exc.$) band
can be constructed according to \cite{Cas94}. The result is:

\begin{equation}
\begin{array}{ll}
|^{100}Mo(Ru), J^+ \sigma \rangle = & | \{2^{\frac {n^N_\pi} 2}\}
(\lambda_\pi ,\mu_\pi );
\ \{2^{\frac {n^N_\nu} 2}\} (\lambda_\nu ,\mu_\nu ); \ 1 (\lambda
\mu )_\sigma K=1 J >_N \\
& | \ (g_{9/2})^{n^A_\pi},  \ J^A_\pi = 0; \
(h_{11/2})^{n^A_\nu},
\ J^A_\nu = 0 >_A \ \ ,  \label{wave}
\end{array}
\end{equation}

Each $(\lambda 's,\mu 's)$ are also depicted in Table 1.
In this approach, instead of assuming the first excited $0^+$ and
the second $2^+$ states as parts of a two phonon triplet, together with
the first $4^+$ states, it is assumed that the first $0^+, 2^+, 4^+$
states of \ru ~are the low energy sector of a rotational band based in the
normal $(\lambda ,\mu )_{g.s.} = (8,6)$ strong coupled pseudo SU(3)
configuration and that the second $0^+, 2^+$ excited
states belong to a second rotational band  described by the $(\lambda
,\mu )_{exc.} = (10,2)$
 pseudo SU(3) configuration.
Note that in the sph-o case the normal parity orbitals ($\tilde N = 3$)
are filled, giving rise to the irrep $(0,0)$. It implies that these
protons are
dynamically inert. It also has the consequence that there exists only one
strong coupled stated $(0,0)_\pi \times  (12,0)_\nu = (12,0)$, and in this
approximation there is not a $0^+$ excited state in our restrictive
Hilbert space.

In order to analyse the spectra and transitions amplitudes of \ru ~we
have
selected a simplified version of the pseudo SU(3)
Hamiltonian\cite{Dra84}, i.e.

\begin{equation}
H = \sum_{\alpha}{H_\alpha - {1 \over 2} \chi \bf{Q}^a \cdot \bf{Q}^a}
   + a L^2 \hspace{2cm} \alpha=\pi ,\nu ,
\end{equation}

\noindent
where  the algebraic quadrupole operator
${\bf {Q}^a}= \sum_{s}{\{{q_\pi}_s  + {q_\nu}_s\}}$
acts only within a shell and do not mix different shells and

\begin{equation}
 H_\alpha = \sum_{s}{\hbar \omega \left\{ \eta_{\alpha s}
	+ {3 \over 2}  -2 k_\alpha \
	  \vec{\bf L}_{\alpha s} \cdot \vec{\bf S}_{\alpha s} - k_\alpha
		\mu_\alpha {L}^{2}_{\alpha s} \right\} - {\rm V_\alpha}}
\approx  \sum_{s}{\hbar \omega \left\{ \tilde\eta_{\alpha s}\right\}
	- {\rm V_\alpha}}
\end{equation}

\noindent
where $\eta_\alpha  = \tilde\eta_\alpha + 1$ denotes the phonon number
operator, $\hbar \omega$
determines the size of the shell and $\alpha = \pi \hbox{ or } \nu$.
 A
constant term ${\rm V_\nu} \ ({\rm V_\pi})$ is included,
which represents the depth of the neutron (proton) potential well.

The abnormal parity sector of the wave function contributes with a
constant to the eigenvalues of this Hamiltonian, because the seniority zero
approximation inhibits any dependence with the total angular
momentum. Thus, the expectation value of the Hamiltonian between  the \ru ~
wave functions only depends on the $(\lambda, \mu )$ and $J=L$ of the
normal parity orbitals. Explicitly:

\begin{equation}
\begin{array}{l}
<^{100}Ru, (\lambda ,\mu ) K=1, L=J~M | H |^{100}Ru, (\lambda ,\mu ) K=1,
L=J~M  > = \\
\hspace{2cm} N_s \hbar \omega - 2 \chi C_2 - ( {\frac 3 2} \chi + a) L(L+1)
\end{array}
\end{equation}

In the above expression $N_s$ represents the number of proton plus
neutron phonons, which depends only on the normal parity occupations
$n^N_\pi$ and $n^N_\nu$ and do not depend on the irrep or on the angular
momentum $L$. The eigenvalues of the  second order
Casimir operator of SU(3) is$C_2 = (\lambda +
\mu + 3) (\lambda + \mu) - \lambda \mu $.
The irrep $(8,6)$ has the maximum value of $C_2 = 190$ and yields
the ground state band. The
second $0^+$  and $2^+$ states were associated with the irrep $(10,2)$, with
$C_2 = 160$.

The two parameters of the Hamiltonian (3) are fitted by means of  the
excitation
energies of the first $2^+$ and the first excited $0^+$ states. The
results are $\chi = 18.83 keV,~ a = 61.75 keV$, which are reasonable numbers
as an interpolation
between those needed to reproduce the rotational spectra of
$^{24}Mg$\cite{Cas89} and $^{238}U$\cite{Cas91}. In the comparison it
must be taken into account that the Refs. \cite{Cas89} and \cite{Cas91} use
symplectic extensions
of SU(3) model and thus they need
smaller coupling constants.

In Table 2 the experimental and theoretical low lying states in \ru ~are
presented, with the angular momentum, parity and energy
stated in each level, as well as the corresponding irrep $(\lambda ,\mu
)$ of each
band. As expected, the energies of the first $4^+$ and the second $2^+$
states
are overestimated. This is a remainder that the present exercise is
a rotational limit of a triaxial nucleus, whose low energy spectra evokes
an anharmonic oscillator, as it was mentioned above.

The {\em average} quadrupolar moment $Q_0$ is related with a particular B(E2)
value as follows:
\begin{equation}
B(E2,0^+\rightarrow 2^+) = {\frac 5 {16 \pi}} |Q_0|^2
\end{equation}

In the pseudo SU(3) scheme, this intrinsic quadrupolar moment is easily
evaluated as\cite{Hir94b}

\begin{equation}
Q_0 = e^{eff}_\pi Q_\pi + e^{eff}_\nu Q_\nu
\hspace{1.5cm} Q_\alpha = {\frac {\eta_\alpha +1}
{\eta_\alpha}} ( 2 \lambda_\alpha +
\mu_\alpha)
 \end{equation}

\noindent where $Q_0$ is given in units of $e~b_0^2$ with
$b_0^2 = {\frac \hbar {M \omega}} = 4.70 fm^2$.
Using the wave functions of the ground state band
(\ref{wave}) the intrinsic quadrupole moments are:

\begin{equation}
Q_\pi = 5.33\hspace{2cm} Q_\nu = 22.5
 \end{equation}

The experimental value is $Q_0(exp) = 57.7 e~b_0^2$. To reproduce this
result unreasonable effective charges $e^{eff}_\pi \approx 2.9 e$ and
$e^{eff}_\nu \approx 1.9 e$ would be needed. This failure of the pseudo
SU(3) model in reproducing the quadrupole moment of \ru ~is related with
the use of
the seniority zero approximation for nucleons in abnormal parity
orbitals. In the case of \ru ~case
 even in the def-o limit there are only two
normal holes and the associated irrep $(0,4)$ gives a very little proton
quadrupole moment. The greater part of the proton quadrupole moment must
come from protons
in the abnormal parity orbital $g_{9/2}$, whose contribution is
neglected in the seniority zero approximation. These limitations were
discussed previously\cite{Bha92,Hir94b}. In this picture large
effective charges  are always
 needed in order to compensate this effect. For the case
of \ru ~these effective charges are
too large. Although we have
presented here some limitations of the rotational model
 of \ru , it is successful in describing the \bb decay of \mo .

We turn now to the study of the two neutrino mode of the double beta decay
\bb of
 \mo  ~into the ground state, the first $2^+$ and the first excited $0^+$
states
 of \ru . The mathematical expressions needed to evaluate the \bb to the ground
 state of \ru ~can be found in \cite{Cas94}. The same formulae works for the
 decay to the first excited $0^+$ state, replacing the strong coupled irrep
$(8,6)$ by $(10,2)$.
The decay to the first $2^+$ state requires a different expression, which can
be found in \cite{Hir94c}. The formulae for this decay resembles that of the
decay to the $0^+$ states, but the energy denominator is elevated to the
third power.
Being in general this energy of the order of $10~MeV$ this power implies a
factor 100 of suppression for this matrix element\cite{Doi85,Gri92,Suh94}.

In Table 3 the dimensionless
 \bb matrix elements for the decay of \mo ~ to the
ground state, the first $2^+$ and the first excited $0^+$ states of \ru ~are
presented. The first two rows  include the results of the
 QRPA calculations obtained  using a
$\delta$-interaction\cite{Gri92} and an interaction derived from the Bonn
one-boson-exchange potential  using G-matrix techniques\cite{Suh94}. The
third row shows the results for the pseudo SU(3) calculation using the
def-o wave functions given in Eq. (\ref{wave}).
The energy denominator used for the decay to the ground state,
 determined by
fixing the energy of the Isobaric Analog  State\cite{Cas94}
 is $E =   E_0 -  2 \hbar \omega k_\pi ( \eta_{\nu} + {1 \over 2}) +
\Delta_C = 7.95 MeV$, where $E_0= {1 \over 2} {\it Q}_{\beta  \beta}+ m_e
c^2$ and $\Delta_C$ is the Coulomb displacement energy.
The matrix elements are given in units of the first or third power of the
inverse electron mass\cite{Doi85}, since the energy denominators have been
divided by the electron rest mass.
For the first $2^+$ state the energy denominator
is equal to $7.68 MeV$ and for the first excited $0^+$
 it is	$7.39 MeV$.  The results
for the sph-o pseudo SU(3) approach are given in the fourth row. As  it
was anticipated in the {\em spherical occupation} limit the decay to
the excited $0^+$ state is absent.

The experimental matrix elements showed in Table 3 were extracted from the
measured half-lives $T^{1/2}_{2\nu}$ using phase space
integrals $G_{GT}$,
which were obtained following the prescriptions given in \cite{Doi88}
with $g_A/g_V = 1.0$. These data are reproduced in the last two rows of
Table 3.

It is remarkable that the sph-o pseudo SU(3) limit reproduces
quite well the experimental matrix element for the \bb to
 the ground state
and that the def-o pseudo SU(3) limit is able to do the same with the
matrix element related with the \bb decay to the first $0^+$ state.
Both the ability to reproduce the numbers and the difficulty to fit both
in the same model are pointing to the description or \ru ~as a
triaxial nucleus.

The \bb matrix element for the $0^+\rightarrow 2^+$ decay is strongly
cancelled in comparison with the QRPA one\cite{Suh94}. In the sph-o case,
which successfully reproduce the $0^+\rightarrow 0^+(g.s.)$ decay, the
reduction has two contributions: i) the matrix element of the
operator $\sigma_1 \cdot \sigma_2$ which connects the $0^+$ states is a
factor ten greater than the matrix element of the operator
$[\sigma_1 \otimes \sigma_2]^2$ which mediates the other decay; ii) the
energy denominator in the pseudo SU(3) model is approximately three times
greater than the energy denominator associated with the first
intermediate state $1^+$, giving a $3^3 = 27$ additional reduction
when compared with the QRPA result.

In order to estimate the effect of the quadrupole-quadrupole correlations
over the \bb matrix elements, we have
performed also a simple estimation of
these matrix elements under the {\em pure seniority zero} approximation.
It is a model in which only the $g^{\pi}_{9/2}$ and $g^{\nu}_{7/2}$
orbitals are
active, with the nucleons paired to angular momentum zero. There are 2
protons and 8 neutrons in \mo  ~and 4 protons and 6 neutrons in \ru .
These two (four) protons are
described in the same way in this simple approach and in the sph-o pseudo
SU(3) scheme but neutrons have strong mixing with their partners in the
$N=4$ oscillator shell in the pseudo SU(3) scheme which is absent in the
other.

The seniority zero matrix elements\cite{Cas94} are exactly the
same as the pure pairing ones\cite{Vog86,Hir90}, {\em i.e.}

\begin{equation}
M_{2\nu}(0^+_i\rightarrow 0^+_f(g.s.)) = \sum\limits_{pn} {{u_p v_n \bar u_n
\bar v_p <p\|\sigma\|n>^2}\over{e_p+e_n-E_0}}
\end{equation}

\noindent
There exist only one $ (p,n)~ =~ (g^{\pi}_{9/2},~ g^{\nu}_{7/2})  $
configuration in this approach, limiting
the sum to only one term. The pairing occupation numbers $v_\alpha^2$ in
the initial (unbarred) and final (barred) nuclei are
$u_p^2 = 0.8, \bar v_p^2 = 0.4, v_n^2 = 1.0, \bar u_n^2 = 0.25$.
Using this numbers together with
the energy denominator given above  we have
obtained the matrix element
for the \bb in this seniority zero approach which is given in the fifth
row of Table 3. The comparison with the pseudo SU(3) matrix elements
exhibits a reduction by a factor two in the sph-o limit and
by a factor three in the def-o one. This reduction is due
to the quadrupole-quadrupole correlations, which strongly mix the
$g^\nu_{7/2}$ with their shell partners, and also mix these neutrons with
the normal parity protons. Using an energy denominator consisting
essentially of twice the pairing gap plus half the Q-value ($\approx 3
MeV$) would increase the seniority zero matrix element making it worst
when compared with the experimental value.

In the present paper we have studied the \bb
of \mo  ~to \ru  ~in the context of the pseudo SU(3) model.
It was also mentioned that in the geometrical model,
while the \ru  ~ground state essentially feels an
anharmonic oscillator potential, the first excited $0^+$ state is
concentrated in the deformed region. We have explored
 the possibility of a
coexistence of shapes, trying to find out a possible explanation for the
difficulties found in previous attempts to describe the \bb of \mo ~ to
excited states of \ru  ~which described the first excited $0^+$ state as a
two phonon state.

We have
 used the pseudo SU(3) model as a deformed limit and
compared it with
the QRPA, which is taken as the spherical limit. Under this scheme, and
with the additional approximation of a seniority zero wave function
 for the
abnormal parity nucleons, we have
generated a rotational spectrum for \ru ~which does not resemble so much the
experimental one. Also, we were not able to generate
enough proton quadrupole moment since we have used
the seniority zero approximation
for abnormal parity nucleons. Besides this limitations, the description
of the \bb was successful. We were able to reproduce the experimental
nuclear matrix elements, without any parameter, but using the {\em
deformed occupation} pseudo SU(3) wave function for \bb to the first excited
$0^+$
state, and the {\em spherical occupation} pseudo SU(3) wave function for the
\bb to
the ground state. Fitting both decays in the same model
is not possible, at least within the present scenario.

\vskip .5cm
We acknowledge P. Vogel for pointing out an incorrect rescaling in the
phase space integrals and S. Pittel for
interesting comments about the microscopic
origin of deformation in \mo ~ and about the seniority
zero approach.
This work was
supported in part by Conacyt under contract 3513-E9310, and by a
Conacyt-CONICET agreement under the project {\em Double Beta Decay.}

\newpage
\vskip 1cm
\centerline{\bf Table captions}

\vskip 2cm
{\bf Table 1} Occupation numbers and $ (\lambda 's,\mu 's) $ for the
{\em deformed occupation (def-o)} and {\em spherical occupation (sph-o)}
pseudo SU(3) wave function for \mo ~ and \ru .

\vskip 1cm
{\bf Table 2} Experimental and theoretically determined energies (in keV)
for the ground state (g.s) end excited (exc.) band of \ru .

\vskip 1cm
{\bf Table 3} The dimensionless \bb  matrix elements for the decay of \mo ~ to
the ground state, the first $2^+$ and the first excited $0^+$ states of
\ru , evaluated with different models. The experimental matrix elements,
measured half lifes $T^{1/2}_{2\nu}$ and phase space integrals $G^{GT}$
are also listed.

\newpage
\centerline {Table 1}
$$
\begin{array}{llllllllll}
&&\vline ~~n^A_\pi &n^A_\nu &\vline ~~n^N_\pi &n^N_\nu &\vline
{}~~(\lambda_\pi ,\mu_\pi )  &(\lambda_\nu ,\mu_\nu ) &(\lambda ,\mu
)_{g.s.} &(\lambda ,\mu )_{exc.} \\
\hline&&\vline &&\vline &&\vline \\
&Mo &\vline ~~4 & 2 &\vline ~~10 &6 &\vline ~~(0,4) &(12,0) &(12,4) \\
def-o &&\vline &&\vline &&\vline \\
&Ru &\vline ~~6 &2 &\vline ~~10 &4 &\vline ~~(0,4) &(8,2) &(8,6) & (10,2) \\
\hline&&\vline &&\vline &&\vline \\
&Mo &\vline ~~2 & 0 &\vline ~~12 &8 &\vline ~~(0,0) &(10,4) &(10,4) \\
sph-o &&\vline &&\vline &&\vline \\
&Ru &\vline ~~4 &0 &\vline ~~12 &6 &\vline ~~(0,0) &(12,0) &(12,0)  \\
\end{array}
$$
\vskip 2cm
\centerline {Table 2}
$$
\begin{array}{clrlr}
&\vline ~~~~~g.s. &band~~~~~~ &\vline ~~~~~exc. &band~~~~~~~ \\
J^+ &\vline ~~~~~th.  &exp.~ &\vline ~~~~~th.  &exp~\\
\hline &\vline & &\vline \\
irrep &\vline ~~(8,6) &&\vline~~~(10,2) &\\
0^+ &\vline ~~~~~~~0 &0~ &\vline ~~~1130 &1130\\
2^+ &\vline ~~~~540 &540~ &\vline ~~~1362 &1670 \\
4^+ &\vline ~~1800 & 1227~ &\vline\\
&\vline & &\vline \\ \hline &\vline & &\vline \\
\end{array}
$$

\newpage
\centerline {Table 3}
$$
\begin{array}{llcc}
\hline
&\vline\\
&\vline~~ 0^+\rightarrow 0^+(g.s.)  &0^+\rightarrow 0^+(exc.)
&0^+\rightarrow 2^+\\ &\vline \\ \hline  &\vline \\
QRPA \cite{Gri92} &\vline ~~~~-0.256 & -0.256 \\  &\vline \\
QRPA \cite{Suh94} &\vline ~~~~~~0.197 & -0.271 & -0.033 \\  &\vline \\
pseudo SU(3) (def-o) &\vline ~~~~-0.108 &~ 0.098 & 1.53 \times 10^{-4}\\
&\vline \\
pseudo SU(3) (sph-o) &\vline ~~~~~~0.152 & & 7.3 \times 10^{-5} \\
&\vline \\ seniority ~0 &\vline ~~~~-0.323 \\
&\vline \\ \hline  &\vline\\
experiment &\vline ~~~~\pm 0.150 &\pm 0.092 & <0.106 \\
&\vline \\ \hline &\vline\\
T^{1/2}_{2\nu} ~[10^{19}yr] &\vline ~~~1.15\cite{Eji91,Moe93}
&178\cite{Bar91}&>11.5 \cite{Kud92}\\&\vline \\
G_{GT}~ [10^{-20} yr^{-1}] &\vline ~~~~~~387 & 6.61  & 76.6 \\
&\vline \\
\end{array}
$$
\end{document}